# Analytical Realization of Finite-Size Scaling for Anderson Localization. Does the Band of Critical States Exist for $d > 2$?

I. M. Suslov

*Kapitza Institute for Physical Problems, Russian Academy of Sciences, Moscow, 119334 Russia*
*e-mail: suslov@kapitza.ras.ru*

**Abstract**—An analytical realization is suggested for the finite-size scaling algorithm based on the consideration of auxiliary quasi-1D systems. Comparison of the obtained analytical results with the results of numerical calculations indicates that the Anderson transition point splits into the band of critical states. This conclusion is supported by direct numerical evidence (Edwards, Thouless, 1972; Last, Thouless, 1974; Schreiber, 1985). The possibility of restoring the conventional picture still exists but requires a radical reinterpretation of the raw numerical data.

## 1. INTRODUCTION

In the previous paper [1] we have analyzed the widespread variant of finite-size scaling based on the concept of the minimal Lyapunov exponent for associated quasi-1D systems [2–5]. It was shown for the 2D case, that the minimal Lyapunov exponent does not obey one-parameter scaling and the usual interpretation of the raw numerical data is basically incorrect. Contrary to the statements made in [3–5], the transition of the Kosterlitz-Thouless type between exponential and power-law localization is possible in 2D systems [2].

The consideration in [1] is based on the study of the second moments for a solution of the Cauchy problem for the Schrödinger equation, and in this respect it is close to [6, 7]. However, justification of approach and interpretation of results are essentially different, and in fact we disagree with most of statements made in [6, 7]. In what follows, the results for dimensions $d > 2$ are presented.

Apart a clear demonstration of existence of the Anderson transition for $d > 2$ (which has a certain formal interest though contains no physical novelty), the present approach leads to the conclusion that there is possible "splitting" of the critical point. When the amplitude of disorder $W$ is changed, two critical points $W_c$ and $W_{c0}$ arise: exponential localization takes place for $W > W_{c0}$ and the metallic phase exists for $W < W_c$, while the interval $W_c < W < W_{c0}$ corresponds to the states with a power-law envelope and strong fluctuations on the small length scales; such properties are widely discussed for the states corresponding to the critical point.

Such conclusion is supported by direct numerical evidence but contradicts the accepted theoretical views on the Anderson transition. The possibility to restore the conventional picture still exists but requires a radical reinterpretation of the raw numerical data: one should admit that the accepted values for critical disorder are strongly exaggerated and in fact the transition occurs at essentially smaller disorder.

## 2. PROPERTIES OF A SOLUTION OF THE CAUCHY PROBLEM

We consider the algorithm of finite-size scaling based on consideration of associated quasi-1D systems: for example, instead of an infinite 3D system one should consider a finite system of size $L \times L \times L_z$, where $L_z \longrightarrow \infty$. A solution of the Cauchy problem for the quasi-1D Schrödinger equation with the initial conditions on the left end allows decomposition of the form [1]

$$\psi_n(r_\perp) = A_1(n, r_\perp)e^{\gamma_1 n} + A_2(n, r_\perp)e^{\gamma_2 n} + \ldots + A_m(n, r_\perp)e^{\gamma_m n}, \quad (1)$$

where $\gamma_s$ are the Lyapunov exponents ($\gamma_1 > \gamma_2 > \ldots > \gamma_m > 0$), $n$ is the discrete longitudinal coordinate (in units of the atomic space), $r_\perp$ is the transverse coordinate and $A_s(n, r_\perp)$ are bounded functions. The Lyapunov exponents $\gamma_s$ exist due to Oceledec's theorem [8] and can be calculated by the transfer matrix method [2]. The minimal Lyapunov exponent $\gamma_{\min} \equiv \gamma_m$ can be used to estimate the correlation length $\xi_{1D}$ of the quasi-1$D$ system ($\xi_{1D} \sim 1/\gamma_{\min}$) and introduce the scaling variable $g = \xi_{1D}/L$, which increases with $L$ in a phase with long-range order and decreases in a phase with short-range correlations [1, 2].

The mean value of $\langle \psi_n(r_\perp) \rangle$ does not have a systematic growth inside the allowed energy band [1], while its second moment allows decomposition of the type (1)

$$\langle \psi_n^2(r_\perp) \rangle = B_1(r_\perp)e^{\beta_1 n} + B_2(r_\perp)e^{\beta_2 n} + \ldots + B_m(r_\perp)e^{\beta_m n}, \quad (2)$$

with the same number of positive exponents $\beta_s$. As it was extensively discussed in [1], the exponents $\beta_s$ provide a rigorous upper bound for $\gamma_s$, $\beta_s \geq 2\gamma_s$, while the order of magnitude equality $\beta_s \sim \gamma_s$ holds for a typical physical situation. The latter estimate follows from the relation $b_s \lesssim a_s$ for the parameters $a_s$ and $b_s$ entering the logarithmically normal distribution [1], which is confirmed by existing analytical results for weak [9] and strong [10] disorder and different numerical investigations [11]. Consequently, the study of decomposition (2) makes it possible to obtain qualitative information on the spectrum of exponents $\gamma_s$ and rigorous restrictions on their behavior.

Consider $d$-dimensional Anderson model describing by the discrete Schrödinger equation

$$\psi_{n+1,\mathbf{m}} + \psi_{n-1,\mathbf{m}} + \sum_i \psi_{n,\mathbf{m}+\mathbf{e}_i} + V_{n,\mathbf{m}}\psi_{n,\mathbf{m}} = E\psi_{n,\mathbf{m}}, \quad (3)$$

where we separated the longitudinal coordinate $n$; $\mathbf{m}$ is a vector cite number in the transverse direction, and $\mathbf{e}_i$ are the unit vectors directed from the cite $\mathbf{m}$ to its nearest neighbors in the plane $n = $ const. Introducing the pair correlators

$$x_{\mathbf{m},\mathbf{m'}}(n) \equiv \langle \psi_{n,\mathbf{m}} \psi_{n,\mathbf{m'}} \rangle,$$
$$y_{\mathbf{m},\mathbf{m'}}(n) \equiv \langle \psi_{n,\mathbf{m}} \psi_{n-1,\mathbf{m'}} \rangle, \quad (4)$$
$$z_{\mathbf{m},\mathbf{m'}}(n) \equiv \langle \psi_{n-1,\mathbf{m}} \psi_{n,\mathbf{m'}} \rangle,$$

one can obtain a closed system of the difference equations (see [1] for details)

$$x_{\mathbf{m},\mathbf{m'}}(n+1) = W^2 \delta_{\mathbf{m},\mathbf{m'}} x_{\mathbf{m},\mathbf{m'}}(n)$$
$$+ \sum_{i,j} x_{\mathbf{m}+\mathbf{e}_i,\mathbf{m'}+\mathbf{e}_j}(n) + x_{\mathbf{m},\mathbf{m'}}(n-1)$$
$$+ \sum_i y_{\mathbf{m}+\mathbf{e}_i,\mathbf{m'}}(n) + \sum_j z_{\mathbf{m},\mathbf{m'}+\mathbf{e}_j}(n), \quad (5)$$

$$y_{\mathbf{m},\mathbf{m'}}(n+1) = -\sum_i x_{\mathbf{m}+\mathbf{e}_i,\mathbf{m'}}(n) - z_{\mathbf{m},\mathbf{m'}}(n),$$

$$z_{\mathbf{m},\mathbf{m'}}(n+1) = -\sum_j x_{\mathbf{m},\mathbf{m'}+\mathbf{e}_j}(n) - y_{\mathbf{m},\mathbf{m'}}(n),$$

where we accept that $E = 0$. The cite energies $V_{n,\mathbf{m}}$ are considered to be statistically independent quantities with the first two moments

$$\langle V_{n,\mathbf{m}} \rangle = 0, \quad \langle V_{n,\mathbf{m}} V_{n',\mathbf{m'}} \rangle = W^2 \delta_{nn'} \delta_{\mathbf{m}\mathbf{m'}}. \quad (6)$$

The dependence of the solution on $n$ is exponential,

$$x_{\mathbf{m},\mathbf{m'}}(n) = x_{\mathbf{m},\mathbf{m'}} e^{\beta n}, \quad y_{\mathbf{m},\mathbf{m'}}(n) = y_{\mathbf{m},\mathbf{m'}} e^{\beta n},$$
$$z_{\mathbf{m},\mathbf{m'}}(n) = z_{\mathbf{m},\mathbf{m'}} e^{\beta n}, \quad (7)$$

and after the formal change of variables

$$x_{\mathbf{m},\mathbf{m'}} \equiv \tilde{x}_{\mathbf{m},\mathbf{m'}-\mathbf{m}} \equiv \tilde{x}_{\mathbf{m},\mathbf{l}}, \text{ etc.}, \quad (8)$$

we have, with tildes omitted,

$$(e^\beta - e^{-\beta}) x_{\mathbf{m},\mathbf{l}} = W^2 \delta_{\mathbf{l},0} x_{\mathbf{m},\mathbf{l}}$$
$$+ \sum_{i,j} x_{\mathbf{m}+\mathbf{e}_i,\mathbf{l}+\mathbf{e}_j-\mathbf{e}_i} + \sum_i y_{\mathbf{m}+\mathbf{e}_i,\mathbf{l}-\mathbf{e}_i} + \sum_j z_{\mathbf{m},\mathbf{l}+\mathbf{e}_j}, \quad (9)$$

$$e^\beta y_{\mathbf{m},\mathbf{l}} = -\sum_i x_{\mathbf{m}+\mathbf{e}_i,\mathbf{l}-\mathbf{e}_i} - z_{\mathbf{m},\mathbf{l}},$$

$$e^\beta z_{\mathbf{m},\mathbf{l}} = -\sum_j x_{\mathbf{m},\mathbf{l}+\mathbf{e}_j} - y_{\mathbf{m},\mathbf{l}}.$$

The coefficients are $\mathbf{m}$-independent and the solution is exponential in $\mathbf{m}$,

$$x_{\mathbf{m},\mathbf{l}} = x_\mathbf{l} e^{i\mathbf{p} \cdot \mathbf{m}}, \text{ etc.}, \quad (10)$$

where allowed values of momentum $\mathbf{p} = (p_1, p_2, \ldots, p_{d-1})$ are $2\pi s/L$, $s = 0, 1, \ldots, L-1$ for each component $p_i$ and correspond to the periodic boundary conditions in the transverse direction, $\psi_{n,\mathbf{m}+L\mathbf{e}_i} = \psi_{n,\mathbf{m}}$. Using (10) and excluding $y_{\mathbf{m},\mathbf{l}}$ and $z_{\mathbf{m},\mathbf{l}}$ from the first equation (9), we come to the equation

$$\sum_{i,j} x_{\mathbf{m}+\mathbf{e}_i+\mathbf{e}_j}[2\cosh\beta e^{-i\mathbf{p}\cdot\mathbf{e}_i} - e^{-i\mathbf{p}\cdot(\mathbf{e}_i+\mathbf{e}_j)} - 1]$$
$$+ 2W^2 \sinh\beta \delta_{\mathbf{m},0} x_\mathbf{m} = 4\sinh^2\beta x_\mathbf{m}, \quad (11)$$

describing the point defect in the $(d-1)$-dimensional block of size $L^{d-1}$ with the periodic boundary condi-

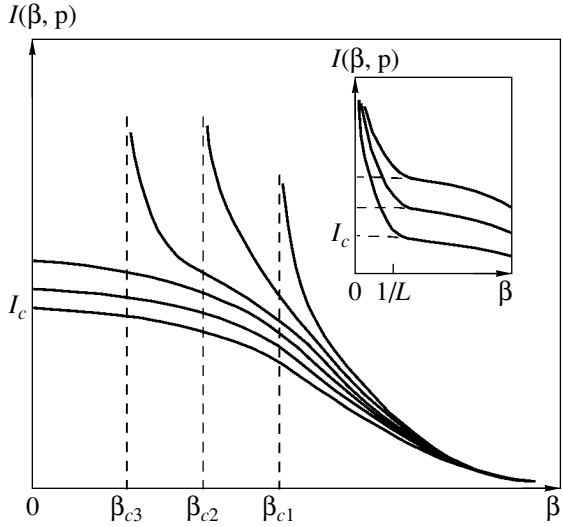

**Fig. 1.** The integral $I(\beta, \mathbf{p})$ as a function of $\beta$ for fixed $\mathbf{p}$ in the continual approximation. Insert shows the same dependences, when discreteness of the sum in (16) is taken into account.

tions $x_{\mathbf{m} + L\mathbf{e}_i} = x_\mathbf{m}$. Equation (11) can be rewritten in the form

$$\sum_{\mathbf{m}'} J_{\mathbf{m}'} x_{\mathbf{m} + \mathbf{m}'} + V\delta_{\mathbf{m}, 0} x_\mathbf{m} = E x_\mathbf{m} \qquad (12)$$

and solved by a standard method [12, 13]. Introducing the Green function $G_{\mathbf{m}, \mathbf{m}'}$ and the spectrum $\epsilon(\mathbf{k})$ of the unperturbed system ($V = 0$)

$$\sum_{\mathbf{m}'}(E\delta_{\mathbf{m}, \mathbf{m}'} - J_{\mathbf{m}' - \mathbf{m}})G_{\mathbf{m}', \mathbf{m}''} = \delta_{\mathbf{m}, \mathbf{m}''}, \qquad (13)$$

$$G_{\mathbf{m}, \mathbf{m}'}(E) = \frac{1}{N}\sum_\mathbf{k} \frac{e^{i\mathbf{k}\cdot(\mathbf{m} - \mathbf{m}')}}{E - \epsilon(\mathbf{k})},$$

$$\epsilon(\mathbf{k}) = \sum_\mathbf{m} J_\mathbf{m} e^{i\mathbf{k}\cdot\mathbf{m}} \qquad (14)$$

($N$ is a total number of sites), one can obtain $x_\mathbf{m} = G_{\mathbf{m}, 0} V x_0$ from Eq. 12 and then the bounded states are determined by the self-consistency equation $1 = VG_{0, 0}$. In the case of (11), this equation takes a form

$$1 = W^2 I(\beta, \mathbf{p}), \qquad (15)$$

$$I(\beta, \mathbf{p}) = \frac{2\sinh\beta}{L^{d-1}}\sum_\mathbf{k}[4\sinh^2\beta + \epsilon_0^2(\mathbf{k}) + \epsilon_0^2(\mathbf{k} - \mathbf{p}) \qquad (16)$$
$$- 2\cosh\beta \epsilon_0(\mathbf{k})\epsilon_0(\mathbf{k} - \mathbf{p})]^{-1},$$

where $\epsilon_0(\mathbf{k})$ is the ordinary tight-binding spectrum with interaction of the nearest neighbors

$$\epsilon_0(\mathbf{k}) = \sum_{i=1}^{d-1} 2\cos k_i. \qquad (17)$$

Summation in (16) is performed over allowed values of momentum $\mathbf{k}$, which are $2\pi s/L$, $s = 0, 1, \ldots, L - 1$ for each component. Equation (15) determines a value of $\beta$ for each of $L^{d-1}$ allowed values of $\mathbf{p}$, and the number of positive exponents $\beta_s$ coincides with the number of positive exponents $\gamma_s$ for the same problem.

For $d = 2$, the expression for $I(\beta, \mathbf{p})$ takes a form

$$I(\beta, \mathbf{p}) = \frac{\sinh\beta}{2(\cosh\beta - \cos p)}$$

$$\times \frac{1}{L}\sum_k \frac{1}{\cosh\beta - \cos(2k - p)} = \frac{\coth(\beta L/2)}{2(\cosh\beta - \cos p)}$$

and corresponds to the results of [1]. The latter equality can be obtained for odd $L$ with the use of the Poisson summation formula.

## 3. INVESTIGATION OF THE SELF-CONSISTENCY EQUATION

Dependences of $I(\beta, \mathbf{p})$ on $\beta$ for fixed $\mathbf{p}$ are shown in Fig. 1. Below we describe the main steps of investigation of Eqs. (15) and (16) leading to this picture.

**Large $\beta$.** In the localized phase, the spectrum of $\beta_s$ does not extend to zero; for fixed $\beta$ and $L \longrightarrow \infty$ one can replace summation in (16) by integration over the first Brillouin zone:

$$I(\beta, \mathbf{p}) = 2\sinh\beta\int\frac{d^{d-1}k}{(2\pi)^{d-1}}[4\sinh^2\beta + \epsilon_0^2(\mathbf{k}) \qquad (18)$$
$$+ \epsilon_0^2(\mathbf{k} - \mathbf{p}) - 2\cosh\beta\epsilon_0(\mathbf{k})\epsilon_0(\mathbf{k} - \mathbf{p})]^{-1}.$$

For large $\beta$, using a specific form of the spectrum $\epsilon_0(\mathbf{k})$, one can obtain

$$I(\beta, \mathbf{p}) = \frac{1}{2\sinh\beta} + \frac{\epsilon_0(\mathbf{p})}{4\sinh^2\beta} + \ldots. \qquad (19)$$

For $\beta \longrightarrow \infty$, all the bunch of curves for different $\mathbf{p}$ in Fig. 1 shrinks in the single curve; it expands, when $\beta$ diminishes. The upper curve of the bunch corresponds to $\mathbf{p} = 0$, while the lower curve corresponds to $\mathbf{p} = \mathbf{G}/2$, where $\mathbf{G}$ is a vector of the reciprocal lattice corresponding to the corner of the Brillouin zone

$$\mathbf{G} = (2\pi, 2\pi, \ldots, 2\pi). \qquad (20)$$

**Upper part of the bunch.** Let us make a shift $\mathbf{k} \longrightarrow \mathbf{k} + \mathbf{p}/2$ in the integral (18). Then the behavior of curves in the upper part of the bunch in Fig. 1 is determined by the region of small **k**, where the integral has a form

$$I(\beta, \mathbf{p}) = 2\sinh\beta \int \frac{d^{d-1}k}{(2\pi)^{d-1}} \frac{1}{\Delta + \sum_{i,j} a_{ij} k_i k_j} \quad (21)$$

with

$$\Delta = 4\sinh^2\beta + [2 - 2\cosh\beta]\epsilon_0^2(\mathbf{p}/2),$$

$$a_{ij} = (4\cosh\beta - 4)\epsilon_0\left(\frac{\mathbf{p}}{2}\right)\cos(p_i/2)\delta_{ij} \quad (22)$$

$$+ 8\cosh\beta \sin\left(\frac{p_i}{2}\right)\sin\left(\frac{p_j}{2}\right).$$

If a vector **p** belongs to the first Brillouin zone ($|p_i| < \pi$), then the quadratic form in the denominator of (21) is positively determined. The quantity $\Delta$ is positive for large $\beta$ and has a form

$$\Delta = \beta^2[4 - \epsilon_0^2(\mathbf{p}/2)] \quad (23)$$

for $\beta \longrightarrow 0$. Under condition $|\epsilon_0(\mathbf{p}/2)| \leq 2$ (which always holds for $d = 2$), the quantity $\Delta$ remains nonnegative for all $\beta$. If **p** is such that $\|\epsilon_0(\mathbf{p}/2)\| > 2$ then $\Delta$ changes a sign at some critical value $\beta_c$.

For $\beta \sim 1$, all eigenvalues of the matrix $a_{ij}$ in (21) are of the order of unity, and the integral has a singularity $\Delta^{(d-3)/2}$ for small $\Delta$ (with the logarithmic branching for odd $d$). As a result, the integral becomes complex for $\Delta < 0$ and equation (15) has no solutions, while the corresponding curve disappears from Fig. 1. For $d \leq 3$, integral (21) diverges at $\Delta \longrightarrow 0$ and the corresponding curve goes to infinity; for $d > 3$, integral (21) is finite, but the discrete sum in (16) still diverges due to a term with **k** = 0. For small $\beta$, the quadratic form in (21) reduces to $(\mathbf{k} \cdot \mathbf{v})^2$, where **v** is a velocity vector with the components $v_i = -2\sin(p_i/2)$, and integral (21) diverges as $\Delta^{-1/2}$ for all $d$.

It is clear from this analysis, that the curves in the upper part of the bunch in Fig. 1, corresponding to sufficiently small **p** (for which $|\epsilon_0(\mathbf{p}/2)| > 2$), one after another go to infinity at points $\beta_{c1}, \beta_{c2}, \beta_{c3}, \ldots$, and only the curves with $|\epsilon_0(\mathbf{p}/2)| < 2$ reach the point $\beta = 0$.

**Lower part of the bunch.** For large $\beta$, the lower curve of the bunch corresponds to **p** = **G**/2 (see (20)), when (18) takes a form

$$I\left(\beta, \frac{\mathbf{G}}{2}\right) = 2\sinh\beta \int \frac{N(\epsilon)d\epsilon}{4\sinh^2\beta + (2 + 2\cosh\beta)\epsilon^2}, \quad (24)$$

where $N(\epsilon)$ is a density of states corresponding to the spectrum $\epsilon_0(\mathbf{k})$. For small $\beta$, one have

$$I\left(\beta, \frac{\mathbf{G}}{2}\right) = \frac{\beta}{2}\int \frac{N(\epsilon)d\epsilon}{\epsilon^2 + \beta^2} \approx \frac{\pi}{2}N(0). \quad (25)$$

For $d \geq 4$, the curve with **p** = **G**/2 remains lowermost for all $\beta$.[1]

For $d = 3$, it is certainly not the case: the two-dimensional tight-binding spectrum $\epsilon_0(\mathbf{k})$ provides the Van Hove singularity $N(\epsilon) \propto \ln(1/|\epsilon|)$ at the band center, and $I(\beta, \mathbf{G}/2)$ diverges as $\ln(1/\beta)$ for $\beta \longrightarrow 0$. As a result, the curve with **p** = **G**/2 does not remain lowermost for small $\beta$.[2]

**Small $\beta$.** Generally, integral (18) is finite for $\beta \longrightarrow 0$, and has a following behavior for small $\beta$:

$$I(\beta, \mathbf{p}) = \begin{cases} I(0, \mathbf{p}) - A(\mathbf{p})\beta^2, & d = 3, \\ I(0, \mathbf{p}) - A(\mathbf{p})\beta, & d \geq 4, \end{cases} \quad (26)$$

i.e. typical curves have a linear or parabolic form. To prove this statement, let us substitute an explicit form of $\epsilon_0(\mathbf{k})$ to (18) and set $p_i = \pi + 2q_i$. Then one have for small $\beta$

$$I(\beta, \mathbf{p}) = \frac{\beta}{2}\int \frac{d^{d-1}k}{(2\pi)^{d-1}} \{(2a_i\cos k_i)^2 \quad (27)$$

$$+ \beta^2[1 + (a_i\cos k_i)^2 - (b_i\sin k_i)^2]\}^{-1},$$

where $a_i = \cos q_i$, $b_i = \sin q_i$ and summation is implied over repeated indices. Consider an integral over one of the component of the vector **k**, e.g. $k_x$. According to (27) it has a structure

$$\beta \int_{-\pi}^{\pi} dk_x \{(\cos k_x - \alpha)^2 \quad (28)$$

$$+ \beta^2(A\cos^2 k_x + B\cos k_x + C\sin k_x + D)\}^{-1}$$

and can be performed by the contour integration. Setting $z = \exp(ik_x)$, we have

$$\beta \int_{|z|=1} \frac{dz}{z} \left\{\left(\frac{z + z^{-1}}{2} - \alpha\right)^2 \right.$$

$$\left. + \beta^2\left[A\frac{(z + z^{-1})^2}{4} + \ldots\right]\right\}^{-1} = \int_{|z|=1} \frac{zdz}{P_4(z)}, \quad (29)$$

---

[1] Tested numerically for $d = 4, 5, 6$.

[2] Ignorance of this fact lead the authors of [6, 7] to zero value of critical disorder for $E = 0$ and $d = 3$.

where $P_4(z)$ is a polynomial in $z$ of degree four, which has two roots $(z_1, z_2)$ inside the circle $|z| < 1$, and two roots $(z_3, z_4)$ outside it. For $\beta \longrightarrow 0$, two pairs of roots merge, while a parametrization

$$z_1 = z_0 - a\beta + b\beta^2, \quad z_2 = z_0^* - a_1\beta + b_1\beta^2,$$
$$z_3 = z_0 + a\beta + b\beta^2, \quad z_4 = z_0^* + a_1\beta + b_1\beta^2 \quad (30)$$

is possible for small $\beta$, if $|\alpha| < 1$. Substitution (30) into (29) shows, that the result is finite for $\beta \longrightarrow 0$ and the correction $O(\beta)$ vanishes. The rule for integration of (28) can be written in the form

$$\beta \int_{-\pi}^{\pi} \frac{dk_x}{(\cos k_x - \alpha)^2 + \beta^2 f^2(\cos k_x, \sin k_x)}$$
$$= \frac{\pi}{\sqrt{1-\alpha^2}} \left[ \frac{1}{f(\alpha, \sqrt{1-\alpha^2})} + \frac{1}{f(\alpha, -\sqrt{1-\alpha^2})} \right] + O(\beta^2). \quad (31)$$

For $|\alpha| > 1$, the first term in the braces in (28) does not turn to zero, and the integrand can be expanded in $\beta$ immediately:

$$\beta \int_{-\pi}^{\pi} \frac{dk_x}{(\cos k_x - \alpha)^2 + \beta^2 f^2(\cos k_x, \sin k_x)}$$
$$= \beta \frac{2\pi\alpha}{(\alpha^2 - 1)^{3/2}} + O(\beta^3). \quad (32)$$

For $d = 3$, the condition $|\alpha| < 1$ can be always provided, performing integration over $k_x$ or over $k_y$ in the first turn; so, the result (31) holds and its structure does not change in the course of integration over remaining variable, in correspondence with (26). It is possible to find explicitly, for which value of $\mathbf{p}$ the integral $I(\beta, \mathbf{p})$ is minimal in the small $\beta$ limit. Taking $\mathbf{p} = (\pi, \pi - 2q)$ and integrating over $k_x$ according to the rule (31), one have

$$I(0, \mathbf{p}) = \frac{1}{8\pi}$$
$$\times \int_{-\pi}^{\pi} \frac{dk_y}{\sqrt{1 - \cos^2 q \cos^2 k_y} \sqrt{1 - \sin^2 q \sin^2 k_y}}. \quad (33)$$

The integral has symmetry in respect to replacement $q$ by $\pi/2 - q$. It diverges for $q \longrightarrow 0$ and $q \longrightarrow \pi/2$, while at $q = \pi/4$ it takes a minimum value ($d = 3$)

$$I_c = \min_{\mathbf{p}} I(0, \mathbf{p}) = \frac{2}{3\pi} K\left(\frac{1}{3}\right) = 0.3432..., \quad (34)$$

where $K(k)$ is the complete elliptic integral. Considering $\mathbf{p} = (\pi - q_x, \pi/2 - q_y)$ with small $q_x$ and $q_y$, one can be convinced that (34) realizes a local minimum over both variables $q_x$ and $q_y$. Numerical investigation shows that this minimum is in fact global.

For $d \geq 4$, the modulus of $\alpha$ in (28) can be greater or smaller than unity, depending on the values of other variables; the result of integration is given by a superposition of (31) and (32), so the term $O(\beta)$ is finite in correspondence with (26). It can be demonstrated explicitly, transforming (27) according to the scheme

$$I(\beta, \mathbf{p}) = \frac{\beta}{2} \int \frac{d^{d-1}k}{(2\pi)^{d-1}} \frac{1}{(2a_i \cos k_i)^2 + \beta^2}$$
$$= \frac{\beta}{2} \int_{-\infty}^{\infty} \frac{d\epsilon}{\epsilon^2 + \beta^2} \int \frac{d^{d-1}k}{(2\pi)^{d-1}} \delta\left(\epsilon - \sum_{i=1}^{d-1} 2a_i \cos k_i\right)$$
$$= \frac{\beta}{2} \int_{-\infty}^{\infty} \frac{d\epsilon}{\epsilon^2 + \beta^2} \int_{-\infty}^{\infty} \frac{dt}{2\pi} \int \frac{d^{d-1}k}{(2\pi)^{d-1}} \quad (35)$$
$$\times \exp\left(it\epsilon - it \sum_{i=1}^{d-1} 2a_i \cos k_i\right)$$
$$= \frac{1}{4} \int_0^{\infty} dt\, e^{-\beta t/2} \prod_i J_0(2a_i t),$$

where $J_0(t)$ is the Bessel function. We have omitted sums in the brackets of (27): the first of them is restricted from above by a quantity of the order of $\beta$, while the second is insignificant for small values $q_i$, which have a main interest. It can be tested, that $I(0, \mathbf{p})$ has a local minimum at $\mathbf{q} = 0$ (i.e. for $\mathbf{p} = \mathbf{G}/2$ in accordance with the previous discussion) with the corresponding value in it

$$I_c = \frac{1}{4} \int_0^{\infty} dt [J_0(t)]^{d-1} = \begin{cases} 0.2241..., & d = 4, \\ 0.2256..., & d = 5, \\ 0.1884..., & d = 6. \end{cases} \quad (36)$$

Expansion of (35) in $\beta$ shows finiteness of the linear correction to $I_c$, in accordance with (26).[3]

**Region of small $\beta$ for finite $L$.** In the previous discussion we considered (16) for finite $\beta$ in the limit $L \longrightarrow \infty$, when summation can be replaced by integration (in fact, it is possible for $\beta \gg 1/L$). The situation is

---

[3] For $a_i = 1$, the logarithmic divergency arises for $d = 5$, so $I(\beta, \mathbf{G}/2) - I(0, \mathbf{G}/2) \sim \beta \ln \beta$.

different, when $L$ is finite and $\beta$ is arbitrary small; then (16) reduces to the form

$$I(\beta, \mathbf{p}) = \frac{2\beta}{L^{d-1}} \sum_{\mathbf{k}} \{[\epsilon_0(\mathbf{k}) - \epsilon_0(\mathbf{k} - \mathbf{p})]^2 \qquad (37)$$
$$+ \beta^2[4 - \epsilon_0(\mathbf{k})\epsilon_0(\mathbf{k} - \mathbf{p})]\}^{-1}.$$

For odd $L$, the difference $\epsilon_0(\mathbf{k}) - \epsilon_0(\mathbf{k} - \mathbf{p})$ is strictly zero for some value $\mathbf{k} = \mathbf{k}^*$, existing for any allowed value of $\mathbf{p}$:[4] it is $\mathbf{k}^* = (\mathbf{p} + \mathbf{g})/2$, where $\mathbf{g}$ is one of vectors of the reciprocal lattice (we take into account, that $\epsilon_0(\mathbf{k}) = \epsilon_0(-\mathbf{k})$, $\epsilon_0(\mathbf{k} + \mathbf{g}) = \epsilon_0(\mathbf{k})$). For $\beta \longrightarrow 0$, the term with $\mathbf{k} = \mathbf{k}^*$ provides the singular contribution, which can be naturally separated:

$$I(\beta, \mathbf{p}) = \frac{2}{\beta L^{d-1}} \frac{1}{4 - \epsilon_0^2(\mathbf{k}^*)} + I_{\text{reg}}(\beta, \mathbf{p}). \qquad (38)$$

Here $\epsilon^2(\mathbf{k}^*) \le \epsilon^2(\mathbf{p}/2)$ for values of $\mathbf{p}$ inside the first Brillouin zone. As a result, the curves in Fig. 1, which have a finite limit for $\beta \longrightarrow 0$ in the continual approximation (they correspond to $4 - \epsilon_0^2(\mathbf{p}/2) > 0$), in fact bend up (due to $4 - \epsilon^2(\mathbf{k}^*) > 0$) and go to infinity (insert in Fig. 1).

## 4. COMPARISON WITH NUMERICAL RESULTS AND GENERAL ANALYSIS OF SITUATION

Matching (26) and (38) for $\beta \sim 1/L$ shows, that the quantity $I_{\text{reg}}(\beta, \mathbf{p})$ is close to the quantity $I(0, \mathbf{p})$, obtained in the continual approximation. The $W$ dependence of the minimal exponent $\beta_{\min}$, entering in (2), is determined by (24), (26) and (38) in the whole region except for the narrow vicinity of the point $W_c = 1/\sqrt{I_c}$ (which diminishes for $L \longrightarrow \infty$). One can see, that $\beta_{\min} \longrightarrow \text{const}$ for $W > W_c$ and $\beta_{\min} \propto 1/L^{d-1}$ for $W < W_c$ in the large $L$ limit. Supposing that the minimal exponents $\beta_{\min}$ and $\gamma_{\min}$ are of the same order of magnitude, one can estimate the correlation length $\xi_{1D}$ of the quasi-1D system as $1/\beta_{\min}$ and introduce the scaling parameter $g = \xi_{1D}/L$. Then its behavior

$$g \sim \begin{cases} L^{d-2}, & W < W_c, \\ 1/L, & W > W_c \end{cases} \qquad (39)$$

indicates the existence of the metallic phase for $W < W_c$ and exponential localization for $W > W_c$ (see discussion in [1]).

However, there is evidence that the relation $\beta_{\min} \sim \gamma_{\min}$ is violated for $d > 2$: according to numerical results [2–4, 14, 15], the quantity $\gamma_{\min}$ turns to zero (for $L = \infty$) in the point $W_{c0}$, different of $W_c$ (Fig. 2a), which is accepted commonly as a point of the Anderson transition.[5] The table lists values of $W_c = 1/\sqrt{I_c}$ following from Eqs. (34) and (35), and values of $W_{c0}$ obtained numerically in the papers [2–4, 14, 15]: these values correspond to the quantity $\tilde{W} = W\sqrt{12}$, because conventionally the cite energies $V_{n,\mathbf{m}}$ are supposed to have a rectangular distribution of width $\tilde{W}$ with $\langle V_{n,\mathbf{m}}^2 \rangle = \tilde{W}^2/12$. Let us discuss the possible interpretations of the arising situation.

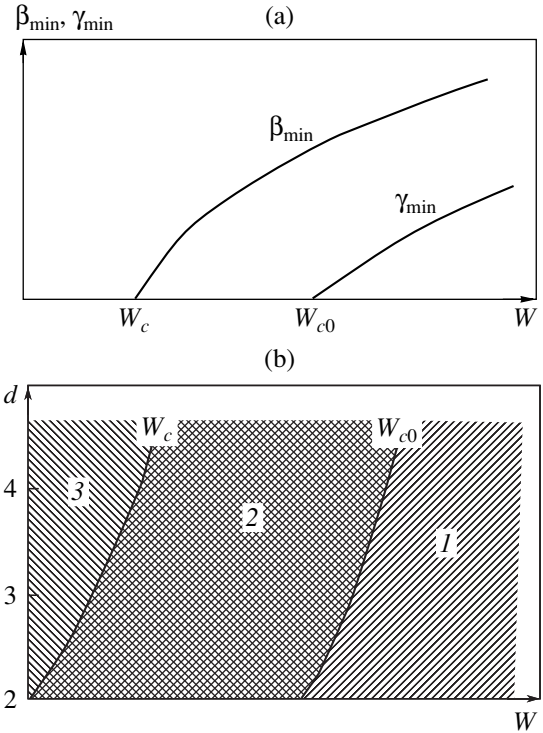

**Fig. 2.** (a) The minimal exponents $\beta_{\min}$ and $\gamma_{\min}$ for $L = \infty$ versus the strength of disorder $W$. (b) Dependences of $W_c$ and $W_{c0}$ on the space dimensionality $d$ in assumption of their continuity. Regions *1*, *2*, *3* correspond to exponential localization, power-law localization and the metallic phase, respectively.

### 4.1. Possibility of the Band of Critical States

There is no doubt, that the region $W < W_c$ corresponds to the metallic phase, while the region $W > W_{c0}$ corresponds to exponential localization. Interpretation of the region $W_c < W < W_{c0}$ is ambiguous.[6] For the sake of simplicity, consider an 1D system, where a solution $\psi_n$ of the Cauchy problem has a following behavior for its typical value and the first two moments

$$\psi_n \sim e^{\gamma n}, \quad \langle \psi_n \rangle \sim 1, \quad \langle \psi_n^2 \rangle \sim e^{\beta n}. \quad (40)$$

---

[4] For even $L$, such value $k^*$ exists not for all $\mathbf{p}$. As a result, a number of positive $\beta_s$ does not coincide with a number of positive $\gamma_s$, and there are difficulties in comparison (1) and (2). For this reason, we do not use even values of $L$.

[5] Numerical results for $d = 5, 6$ were presented by I. Zharekeshev [16] but remained unpublished.

[6] It is clear from the analysis in [1] that $W_{c0}$ and $W_c$ are the points, where the parameters $a$ and $b$ entering the logarithmically normal distribution vanish correspondingly; hence, both of these points have a real physical sense.

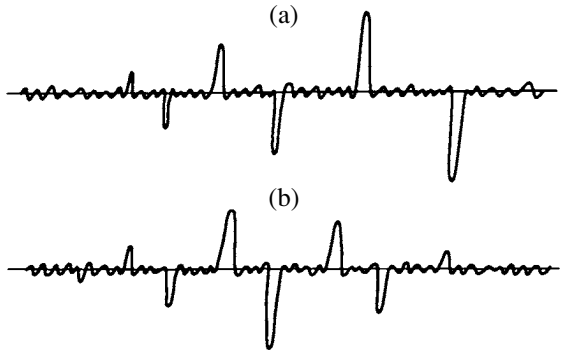

**Fig. 3.** A solution of the Cauchy problem (a) and an eigenfunction of the 1D system (b) in the situation $\gamma = 0$, $\beta > 0$.

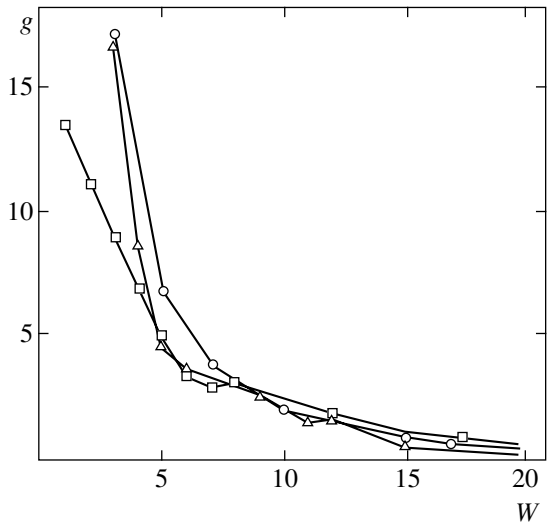

**Fig. 4.** The parameter $g$, characterizing sensitivity to the boundary conditions, versus $W$ for different $L$ [18]: $L = 2$ (□), $L = 4$ (○), $L = 6$ (△).

If $\beta > 0$ and $\gamma = 0$, then $\psi_n$ has no systematic growth but

Values of critical disorder for different dimensions $d$ and the rectangular distribution of $V_{n,\mathbf{m}}$

| $d$ | $W_c$ | $W_{c0}$ |
|---|---|---|
| 3 | 5.91 | 16.5 |
| 4 | 7.32 | 34 |
| 5 | 7.29 | – |
| 6 | 7.98 | – |

contains rare splashes of increasing amplitude (see Fig. 3a). Eigenfunctions of the 1D system can be constructed by matching two solutions of type (40), propagating from different ends of the system: such construction shows existence of the delocalized component, as well as the localized structure of individual splashes (Fig. 3b). The simplest interpretation suggests that an eigenfunction is a hybrid state, i.e. a superposition of the localized and delocalized functions [1].

However, there is another possible interpretation. Zero value for $\gamma$ forbids only the exponential growth for the typical value of $\psi_n$ and does not exclude possibility of its more slower (power-law) increase. As for the envelope of the localized component, its form depends on the statistics of splashes and can be power-like as well as exponential. Indeed, let the splashes occur at points $x_n$, have a width $\Delta_n$ and a random height of the order of $h_n$; then the histogram of the distribution $P(\psi_n)$ consists of rectangles of width $h_n$ and height $\Delta_n/x_n$. In order $\psi_n$ has no systematic growth, its distribution $P(\psi_n)$ should be normalizable, so the quantity $\epsilon_n = h_n \Delta_n / x_n$ should decrease quicker than $1/n$. On the other hand, variance $h_n^3 \Delta_n / x_n \sim \epsilon_n h_n^2$ increases as $e^{\beta n}$ and a power-law envelope $h_n \sim x_n^\alpha$ is possible for

$$x_n \sim (e^{\beta n}/\epsilon_n)^{1/2\alpha}, \quad h_n \sim (e^{\beta n}/\epsilon_n)^{1/2},$$
$$\Delta_n \sim \epsilon_n (e^{\beta n}/\epsilon_n)^{(1-\alpha)/2\alpha}. \quad (41)$$

One can see, that a situation with $\gamma = 0$, $\beta > 0$ may correspond to eigenfunctions with a power-like envelope and strong fluctuations on the small scales; such properties are extensively discussed for the states corresponding to a critical point [17]. Consequently, vanishing of $\beta_{min}$ and $\gamma_{min}$ in the different points (Fig. 2a) may correspond to existence of the whole band of critical states in the interval $W_c < W < W_{c0}$. Such picture matches well with the situation for $d = 2$ discussed in [1]: a value $W_c$ tends to zero in the limit $d \longrightarrow 2$ due to the absence of the metallic state in the 2D case, while $W_{c0}$ remains finite due to existence of the Kosterlitz-Thouless type transition between exponential and power-law localization (Fig. 2b).[7]

Let us come to comparison with numerical results. In spite of the large number of publications and the claims of some authors for high accuracy in determining the transition point, there are only few papers where the Anderson transition is studied directly as a change in the character of the wave functions. In fact, practically all papers since 1981 use one-parameter scaling

---

[7] It should be noted, that power-law localization has different manifestations in finite-size scaling for $d = 2$ and $d > 2$: as $\beta_{min} \sim \gamma_{min} \sim 1/L$ in the first case and as $\beta_{min} \sim 1$, $\gamma_{min} \longrightarrow 0$ (for $L \longrightarrow \infty$) in the second case.

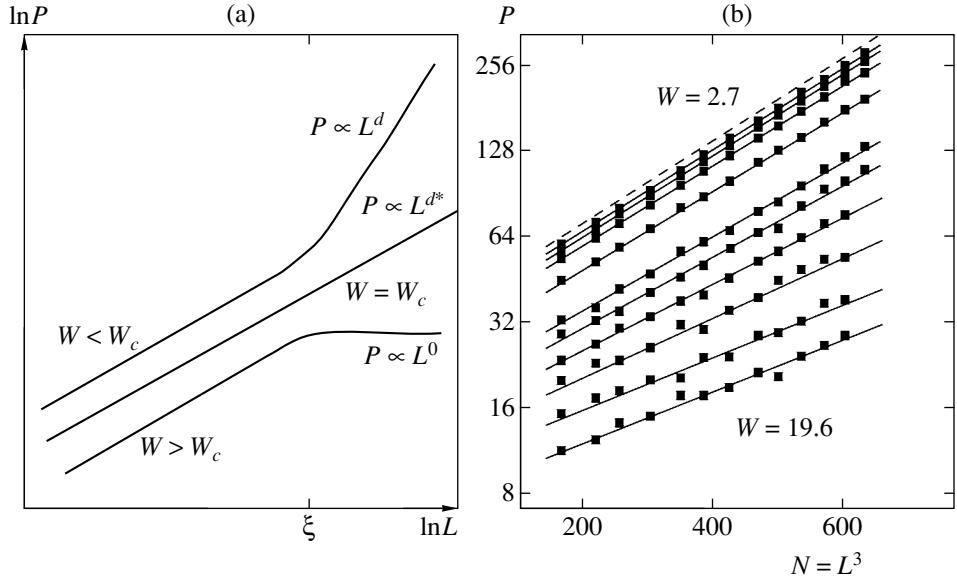

**Fig. 5.** (a) The expected behavior of the participation ratio $P$ versus a system size $L$ for different $W$. (b) Numerical results of the paper [22]: dependences from top to bottom correspond to values $W = 2.7$, $3.9$, $5.5$, $7.8$, $10.9$, $12.3$, $13.9$, $15.6$, $17.5$, and $19.6$ for the 3D Anderson model with the Gaussian disorder. The dashed curve corresponds to the dependence $P \propto L^d$.

and rely on the study of quantities that have rather indirect relation to the Anderson transition.

Existence of the band of unusual states for $5 < W < 15$ (compare with table) in the 3D diamond lattice was noted in the early paper by Edwards and Thouless [18]. The scaling parameter $g$, characterizing sensitivity to the boundary conditions (cited as "Thouless number" or dimensionless conductance [19] in the contemporary literature), was practically independent of $L$ for these states, contrary to the expected growth in the metallic phase and the decrease in the localized phase (Fig. 4). Later, Last and Thouless [20] interpreted these states in terms of power localization.

Existence of the band of critical states is also confirmed by the studies of the participation ratio

$$P = \left(\sum_n |\psi_n|^2\right)^2 \bigg/ \sum_n |\psi_n|^4. \qquad (42)$$

For a finite system in the form of the cube with side $L$, the quantity $P$ behaves as $L^d$ for extended states and as $L^0$ for exponentially localized states. At the critical point, one expects the behavior $P \sim L^{d*}$ in accordance with existence of the fractal dimensionality $L^{d*}$ [21]; the same behavior is expected in the localized and metallic phases on the scales $L \lesssim \xi$ where a system is indistinguishable of the critical one. As a result, the behavior of $P$ in the log–log coordinates should have a form presented in Fig. 5a.

The numerically found behavior [22] is in a striking contrast with Fig. 5a: the power-law dependences $P \propto L^\alpha$ are observed for all $W$, with the exponent $\alpha$ depending on the strength of disorder (Fig. 5b). However, such behavior agrees excellently with existence of the band of critical states. According to table, three upper lines of Fig. 5b (with $W = 2.7$, $3.9$, $5.5$) correspond to the region $W < W_c$ and their slope does not contradict the dependence $P \propto L^d$ with Euclidean dimensionality $d$. The rest of the curves (with $W = 7.8$, $10.9$, $12.3$, $13.9$, $15.6$, $17.5$, and $19.6$) correspond to the region $W_c < W < W_{c0}$ (the Gaussian distribution for $V_{n\mathbf{m}}$ was used, for which $W_{c0} = 21$) and their behavior agrees with the dependence $P \propto L^{d*}$, where the fractal dimensionality $d*$ depends on disorder inside the band of critical states.

*4.2. Possibility of Restoring the Conventional Picture*

If the presented picture is correct, the localization theory appears in a difficult situation: the possibility of the band of critical states is not predicted in any existing variant of theory. The only optimistic outlook is as follows. The exponent $\beta_{\min}$ (and, consequently, the point $W_c$) is determined by only the first two moments of $V_{n\mathbf{m}}$ (see (6)), while the exponent $\gamma_{\min}$ is sensitive to the whole distribution $P(V)$: e.g. values $W_{c0}$ are different for the rectangular and Gaussian distributions. It is possible that $W_{c0}$ can be diminished till $W_c$ by the appropriate choice of $P(V)$ (the possibility $W_{c0} < W_c$ is excluded by inequality $\beta_{\min} \geq 2\gamma_{\min}$ [1]). In this case, the existing theories (e.g. [21, 14, 15]) describe a situation with $W_{c0} = W_c$ and give a zero approximation to the general case:[8] then the nearest aim of the theory should be seen in explanation the mechanism, due to which the isolated critical point is expanded into the band of critical states. The outlined possibility matches well with the situation for $d = 2$ (Fig. 2b) where existence or absence of the Kosterlitz-Thouless type transition depends on a specific model [1].

---

[8] The transition point is not predicted reliably by any existing theory; it is either introduced phenomenologically, or estimated with crude approximations.

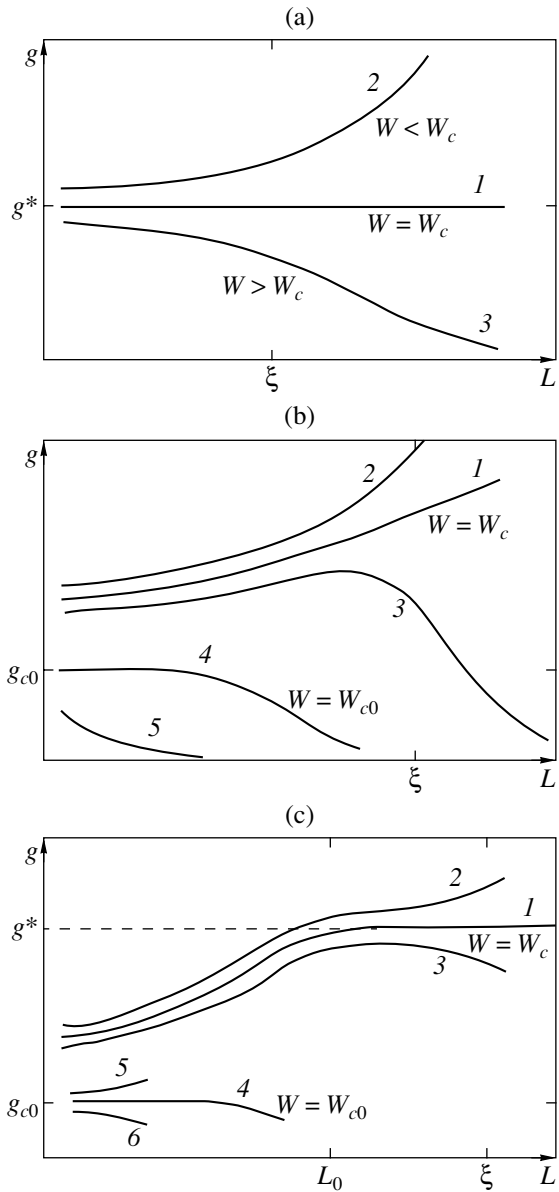

**Fig. 6.** The Thouless parameter $g$ as a function of the length scale $L$ for (a) $d = 3$, (b) $d > 4$, (c) $d = 4 - \epsilon$.

In fact, possibilities of restoring the conventional picture are more wide but require a radical reinterpretation of the numerical data. It should be noted, that numerical calculations do not demonstrate vanishing $\gamma_{min}$ for $W < W_{c0}$ (in the limit $L \longrightarrow \infty$) directly; this conclusion is usually drawn using the interpretation of results in terms of one-parameter scaling (when the parameter $g = 1/\gamma_{min}L$ is supposed to depend only on the ratio $L/\xi$). According to [1], the minimal Lyapunov exponent is a wrong scaling variable and does not obey one-parameter scaling: as a result, the sense of the point $W_{c0}$ becomes somewhat doubtful. In fact, the situation is more complicated: the point $W_{c0}$ is distinguished not only in the studies of the Lyapunov exponents but for many other quantities [25]. However, this fact can also be explained.

The natural scaling variable in the localization theory is the Thouless parameter $g$ [19], which is supposed to obey the Gell-Mann–Low equation

$$\frac{d \ln g}{d \ln L} = \beta(g). \qquad (43)$$

If equation (43) has a fixed point $g^*$ (such as $\beta(g^*) = 0$), then the typical $g(L)$ dependences have a form shown in Fig. 6a: the Thouless parameter $g$ remains constant for the critical point (curve 1) and tends to zero or infinity in the localized and metallic phases correspondingly (curves 2, 3). With increasing of dimensionality $d$, equation (43) is violated, because one of irrelevant scaling parameters (let it be $h$) becomes relevant, as the upper critical dimension $d_{c2}$ is reached [26]. As a result, in the vicinity of $d_{c2}$ one should use the two-parameter scaling equations

$$\frac{d \ln g}{d \ln L} = \beta(g, h), \quad \frac{d \ln h}{d \ln L} = \gamma(g, h), \qquad (44)$$

which can be reduced to a form [26]

$$\begin{aligned}\frac{d \ln g}{d \ln L} &= (d-2) + \tilde{\beta}\left(\frac{g}{h}\right), \\ \frac{d \ln h}{d \ln L} &= (d - d_{c2}) + \frac{b}{h},\end{aligned} \qquad (45)$$

where $d_{c2} = 4$ [27], $b > 0$. Investigation of (45) for $d \geq 4$ shows that the Thouless parameter $g$ does not remain constant in the critical point but increases as $L^{d-4}$ for $d > 4$ and logarithmically for $d = 4$ (Fig. 6b, curve 1). The growth becomes quicker, as $L^{d-2}$, in the metallic phase (curve 2), while the reentrant behavior is realized in the localized phase in the vicinity of the transition (curve 3). In the deep of the localized phase, the parameter $g$ decreases monotonically (curve 5), and by continuity there exists curve 4, corresponding to approximately constant $g$ in the small $L$ region. Value $g_{c0}$, corresponding to the initial part of curve 4, will be interpreted as a critical point, if a formal treatment in terms of one-parameter scaling is performed.

Situation for $d = 4 - \epsilon$ is characterized by existence of the large scale

$$L_0 \propto \exp(\text{const}/\epsilon) \qquad (46)$$

(Fig. 6c), as in the ordinary theory of critical phenomena [28]. For $L \gg L_0$, one-parameter scaling holds in the vicinity of $g^*$ (compare curves 1, 2, 3 in Fig. 6c and Fig. 6a), while fictive one-parameter scaling arises in the vicinity of $g_{c0}$ for $L \ll L_0$ (curves 4, 5, 6). The points $g^*$ and $g_{c0}$ are the roots of the equations $\beta(g, h^*) = 0$ and $\beta(g, h_0) = 0$ correspondingly, where $h^* = b/\epsilon$ is the limiting value of the parameter $h$ for $L \longrightarrow \infty$ (existing according to (45)) and $h_0$ is its initial value, which remains practically unchanged in the $L \ll L_0$ region. Existence of the large scale $L_0$ is also possible for $d = 3$, if *const* in (46) has a value of several units.

It is clear from the previous discussion, that a formal treatment of the dependence $g(L)$ for small $L$ in terms of one-parameter scaling is rather doubtful for $d \geq 4$ and $d = 4 - \epsilon$ (and may be for $d = 3$): it inevitably distinguishes a fictive critical point $g_{c0}$, which will manifest itself in all physical quantities. In fact, the point $g_{c0}$ (which corresponds to the amplitude of disorder $W_{c0}$) lies in the deep of the localized phase. The interval $W_c < W < W_{c0}$ corresponds to the reentrant behavior of the Thouless parameter $g$: it affects the properties of eigenfunctions but does not change their exponentially localized character. From this point of view, difference between $W_{c0}$ and $W_c$ in table should be considered as an artifact, related with not sufficiently large system size for $d = 3$ and the principal inapplicability of one-parameter scaling for $d \geq 4$. Correspondingly, the results presented in Fig. 4 and Fig. 5b do not have a deep sense and reflect a transient behavior related with relaxation of the parameter $h$ to its limiting value $h^*$.

It should be noted, that incorrect determination of the transition point ($g_{c0}$ instead $g^*$) leads to the incorrect value of the critical exponent $\nu$ of the correlation length: for $d < 4$, it will be determined by the derivative $\beta'_g(g_{c0}, h_0)$ instead of $\beta'_g(g^*, h^*)$ as it should be. May be, it will resolve the contradictions between analytical and numerical results that was discussed by the author in [1] and [29].

## 5. CONCLUSIONS

The above analysis reveals the serious contradictions between the existing theoretical views and the available numerical results. In the framework of the presented picture, practically all numerical data for $d > 2$ indicate the existence of the whole band of critical states, which is not described by existing theories. In particular, it cannot be explained by the one-parameter scaling theory [21], and the conventional interpretation for the majority of numerical data becomes self-contradictory.

In order to clarify the situation, it is desirable to extend the studies of the participation ratio (Fig. 5b) to the region of essentially greater $L$. It is technically possible, because such calculations are performed presently for the systems, whose size is 5–10 times greater, but unfortunately only for disorder corresponding to the point $W_{c0}$ [30]. As a result of such studies, the picture in Fig. 5b either remain unchanged, or begin to change into direction of Fig. 5a. In the former case, existence of the band of critical states will be essentially confirmed. In the latter case, such studies will inevitably reveal the large scale $L_0$, which was discussed in Section 4. In any case, reinterpretation of practically all numerical data appears to be inevitable.

This work is partially supported by the Russian Foundation for Basic Research (project no. 03-02-17519).